\documentclass{ws-procs9x6}

\newcommand{\beq}{\begin{equation}}
\newcommand{\eeq}{\end{equation}}
\newcommand{\beqs}{\begin{eqnarray}}
\newcommand{\eeqs}{\end{eqnarray}}
\newcommand{\lsim}{\mathrel{\raisebox{-
.6ex}{$\stackrel{\textstyle<}{\sim}$}}}

\begin{document}

\title{Some Recent Results on Models of Dynamical Electroweak Symmetry
  Breaking} 

\author{R. Shrock$^*$}

\address{C. N. Yang Institute for Theoretical Physics\\
State University of New York\\
Stony Brook, New York 11794, USA\\
$^*$email: robert.shrock@sunysb.edu}

\begin{abstract}
We review some recent results on models of dynamical electroweak symmetry
breaking involving extended technicolor. 

\end{abstract}

\bodymatter

\section{Introduction}

The origin of electroweak symmetry breaking (EWSB) is an outstanding unsolved
question in particle physics. In the Standard Model with gauge group $G_{SM} =
{\rm SU}(3)_c \times {\rm SU}(2)_L \times {\rm U}(1)_Y$, this symmetry breaking
is produced by postulating a Lorentz scalar Higgs field $\phi = {\phi^+ \choose
\phi^0}$ with weak isospin $I=1/2$ and weak hypercharge $Y=1$ and assuming that
its potential, $V=\mu^2 \phi^\dagger \phi + \lambda (\phi^\dagger \phi)^2$, has
$\mu^2 < 0$, thereby leading to a nonzero vacuum expection value for $\phi$.
However, this mechanism is unsatisfying for several reasons: (i) the EWSB is
put in by hand and no explanation is provided as to why $\mu^2$ is negative
when, {\it a priori}, it could be positive; (ii) $\mu^2$ and hence $m_H^2 =
-2\mu^2 = 2\lambda v^2$ are unstable to large radiative corrections from much
higher energy scales (the gauge hierarchy problem), so that extreme fine-tuning
is needed to keep the Higgs mass of order the electroweak scale; (iii) the SM
accomodates, but does not explain, fermion masses and mixing, via Yukawa
couplings of the generic form (suppressing the matrix structure) $m_f \simeq
y_f v/\sqrt{2}$; the $y_f$ values and generational hierarchy are put in by hand
with some $y_f$'s ranging down to $10^{-5}$ with no explanation.

  These facts have motivated an alternative approach based on dynamical
electroweak symmetry breaking driven by a strongly coupled vectorial gauge
interaction, associated with an exact gauge symmetry, called technicolor (TC)
\cite{tc}.  The EWSB is produced by the formation of bilinear condensates of
technifermions.  To communicate this symmetry breaking to the SM fermions
(which are technisinglet), one embeds technicolor in a larger, extended
technicolor (ETC) theory \cite{etc,etcrev}.  In this talk we will review some
of our recent results in this area \cite{nt}-\cite{sg} obtained in
collaboration with T. Appelquist, M. Piai, N. Christensen, and M. Kurachi. (The
work with Kurachi is also discussed in his talk \cite{mk}.)

As further motivation, one may recall that in both of two major previous cases
where fundamental scalar fields were used to model spontaneous symmetry
breaking, the actual underlying physics did not involve fundamental scalar
fields but instead a bilinear fermion condensate.  The first of these was
superconductivity, where the phenomenological Landau-Ginzburg model made use of
a complex scalar field $\phi = \rho e^{i\theta}$ with a free energy functional
of the form $V= c_2|\phi|^2 + c_4|\phi|^4$, where $c_2 \propto (T - T_c)$ as $T
\to T_c$ (with $T_c$ being the critical temperature).  Hence, for $T < T_c$,
$c_2 < 0$ and $\langle \phi \rangle \ne 0$.  However, the true origin of
superconductivity is the dynamical formation of a condensate of Cooper pairs,
$\langle e e \rangle$.  A second example is spontaneous chiral symmetry
breaking in hadronic physics.  In the Gell-Mann L\'evy $\sigma$ model for this
phenomenon, the chiral symmetry breaking is a result of the coefficient of the
quadratic term in the potential being arbitrarily chosen to be negative,
producing a nonzero vev of the $\sigma$ field, quite analogous to the Higgs
mechanism.  However, the real origin of spontaneous chiral symmetry breaking in
quantum chromodynamics (QCD) is the dynamical formation of a bilinear quark
condensate $\langle \bar q q \rangle$.  Perhaps these previous examples might
serve as a guide for thinking about the physics underlying EWSB.

Actually, one already knows of a source of dynamical electroweak symmetry
breaking, namely QCD.  Consider, for simplicity, QCD with $N_f=2$ quarks, $u$,
$d$, taken to be massless. The quark condensate $\langle \bar q q \rangle =
\langle \bar q_L q_R \rangle + \langle \bar q_R q_L \rangle$, transforms as
$I_w = 1/2$, $|Y|=1$.  By itself this theory would produce nonzero $W$ and $Z$
masses $m_W^2 \simeq (g^2/4) f_\pi^2$, \ $m_Z^2 \simeq (1/4)(g^2+g'^2)f_\pi^2$.
With $f_\pi = 92$ MeV, this yields $m_W \simeq 30$ MeV, $m_Z \simeq 34$
MeV. These $W$ and $Z$ masses satisfy the tree-level relation $\rho=1$, where
$\rho = m_W^2/(m_Z^2 \cos^2\theta_W)$. While the scale here is too small by $
\sim 10^3$ to explain the observed $W$ and $Z$ masses, it suggests how to
construct a model with dynamical EWSB.

  We shall consider a technicolor theory with gauge group ${\rm SU}(N_{TC})$
and a set of technifermions, generically denoted $F$, with zero Lagrangian
masses, transforming according to the fundamental representation of the
group. The scale of confinement and chiral symmetry breaking in this theory is
denoted $\Lambda_{TC}$ and is of order the electroweak scale.  One assigns the
SU(2)$_L$ representations of the technifermions so that their left and right
components form SU(2)$_L$ doublets and singlets, respectively.  A minimal
choice is the ``one-doublet'' model, with ${F_u^\tau \choose F_d^\tau}_L$ and
$F_{uR}^\tau$, $F_{dR}^\tau$, where $\tau$ is a TC gauge index and $Y=0$
($Y=\pm 1$) for the SU(2)$_L$ doublet (singlets).  Since the technicolor theory
is asymptotically free, it follows that, as the energy scale decreases,
$\alpha_{TC}$ increases, eventually producing condensates $\langle \bar F_u F_u
\rangle$ and $\langle \bar F_d F_d \rangle$ transforming as $I_w=1/2$, $|Y|=1$,
breaking EW sym. at $\Lambda_{TC}$.  It follows that, to leading order, $m_W^2
= g^2f_{TC}^2 N_D/4$ and $m_Z^2 = (g^2 + g'^2) f_{TC}^2 N_D/4$, where $f_{TC}
\lsim \Lambda_{TC}$ is the TC analogue of $f_\pi \lsim \Lambda_{QCD}$ and $N_D$
is the number of SU(2)$_L$ technidoublets ($N_D=1$ in this case).  These masses
satisfy the tree-level relation $\rho=1$ \cite{ssvz}.  For this minimal
example, $f_{TC} \simeq 250$ GeV.  Another class of TC models uses one SM
family of technifermions \cite{onefamily},
\beq
{U^{a \tau} \choose D^{a \tau}}_L \ , \quad \quad  U^{a \tau}_R, \ \ 
D^{a \tau}_R
\eeq
\beq
{N^{\tau} \choose E^{\tau}}_L \ , \quad \quad  N^{\tau}_R, \ \ E^{\tau}_R
\eeq
(where $a$ and $\tau$ are color and technicolor indices, respectively) 
with the usual $Y$ assignments.  Again, there is technifermion condensate
formation, with approximately equal condensates $\langle \bar F F \rangle$ for
$F=U^a, \ D^a, \ N, \ E$, generating dynamical technifermion masses
$\Sigma_{TC} \sim \Lambda_{TC}$. In this class of models $N_D=N_c+1=4$, so 
$f_{TC} \simeq 125$ GeV.

Technicolor has the potential to solve/explain various problematic and/or
mysterious features of the Standard Model: (i) given the asymptotic freedom of
the TC theory, the condensate formation and hence EWSB are automatic and do not
require any {\it ad hoc} parameter choice like $\mu^2 < 0$ in the SM; (ii)
because TC has no fundamental scalar field, there is no hierarchy problem;
(iii) because $\langle \bar F F \rangle = \langle \bar F_L F_R \rangle +
\langle \bar F_R F_L \rangle$, technicolor explains why the chiral part of
$G_{SM}$ is broken and the residual exact gauge symmetry, ${\rm SU}(3)_c \times
{\rm U}(1)_{em}$, is vectorial.  The fact that this latter symmetry is
vectorial is, of course, crucial in allowing nonzero mass terms for fermions
that are nonsinglets under color or charge.

 In order to give masses to quarks and leptons, one must communicate the EWSB
in the technicolor sector to these SM fermions, and hence, as mentioned above,
one must embed technicolor in the larger ETC theory, with ETC gauge bosons
$V^i_\tau$ transforming (technisinglet) SM fermions into technifermions and
vice versa.  To satisfy constraints on flavor-changing neutral current (FCNC)
processes, the ETC gauge bosons must have large masses.  These masses can arise
from self-breaking (tumbling) of the ETC chiral gauge symmetry.  The ETC theory
is arranged to be asymptotically free, so as the energy decreases from a high
scale, the ETC coupling $\alpha_{_{ETC}}$ grows and eventually becomes large
enough to form condensates which sequentially break the ETC symmetry group down
at scales $\Lambda_j$, $j=1,2,3$ to a residual exact TC subgroup, where 
\beq
\Lambda_1 \simeq 10^3 \ {\rm TeV}, \quad\quad
\Lambda_2 \simeq 50-100 \ {\rm TeV}, \quad\quad
\Lambda_3 \simeq {\rm few \ TeV},
\label{lambdavalues}
\eeq
We will mainly focus on SU($N_{ETC}$) ETC models with one-family TC. These
gauge the Standard Model fermion generation index and combine it with the TC
index $\tau$, so
\beq
N_{ETC}=N_{gen.} + N_{TC} \ = \ 3 + N_{TC} \ . 
\label{nrel}
\eeq

To be viable, modern TC models are designed to have a coupling $g_{_{TC}}$ that
gets large, but runs slowly (``walks'') over an extended interval of energy
(WTC) \cite{wtc1}-\cite{chiv}.  For sufficiently many technifermions, the TC
beta function has a second zero (approximate infrared fixed point of the
renormalization group) at a certain $\alpha_{_{TC}}=\alpha_{_{IR}} \ne 0$. As
the number of technifermions, $N_f$, increases, $\alpha_{_{IR}}$ decreases. In
WTC, one arranges so that $\alpha_{_{IR}}$ is slightly greater than the
critical value, $\alpha_{cr}$, for $\langle \bar F F \rangle$ formation.  As
$N_f \nearrow N_{f,cr}$, $\alpha_{_{IR}} \searrow \alpha_{cr}$.  Combining the
calculation of $\alpha_{_{IR}}$ from the two-loop beta function and an estimate
of $\alpha_{cr}$ from the Schwinger-Dyson equation, one finds $N_{f,cr} =
2N_{TC}(50N_{TC}^2-33)/[5(5N_{TC}^2-3)]$ \cite{chipt2}.  Hence, for $N_{TC}=2$,
one has $N_{f,cr} \simeq 8$.  Thus, if $N_{TC}=2$, a one-family TC theory, with
its $N_f=N_w(N_c+1)=8$ technifermions plausibly exhibits walking behavior.

In a walking TC theory, as energy scale decreases, $\alpha_{_{TC}}$ grows, but
its rate of increase, $|\beta|$, decreases toward zero as $\alpha_{_{TC}}
\nearrow \alpha_{_{IR}}$, where $\beta=0$.  Eventually, $\alpha_{_{TC}}$
exceeds $\alpha_{cr}$, $\langle \bar F F \rangle$ forms, and the technifermions
gain dynamical masses $\Sigma_{TC} \simeq \Lambda_{TC}$.  WTC has several
advantages: (i) SM fermion masses are enhanced by the factor $\eta = \exp \big
[ \int_{\Lambda_{TC}}^{\Lambda_w} \, \mu^{-1} d\mu \ 
\gamma(\alpha_{_{TC}}(\mu)) \big ]$; (ii) hence, one can increase ETC scales
$\Lambda_i$ for a fixed $m_{f_i}$, reducing FCNC effects; (iii) $\eta$ also
enhances masses of pseudo-Nambu Goldstone bosons (PNGB's); and (iv) the walking
can reduce the value of the $S$ parameter \cite{pt}, given by
\beq
\frac{\alpha_{em}S}{\sin^2(2\theta_W)} = \frac{\Pi_{ZZ}(m_Z^2)-\Pi_{ZZ}(0)}
{m_Z^2} \ . 
\label{s}
\eeq
The value $N_{TC}=2$ is also motivated by the fact that it makes possible a
mechanism to explain light neutrinos in (E)TC \cite{nt,lrs}.  Substituting this
value $N_{TC}=2$ into eq. (\ref{nrel}) yields $N_{ETC}=5$.

\section{ETC Models}

TC/ETC models are very ambitious, since a successful model would explain not
only EWSB but also the spectrum of Standard Model fermion masses. Thus it is
perhaps not surprising that no fully realistic model of this type has been
constructed yet.  These models are subject to several strong constraints from
neutral flavor-changing current processes and precision electroweak data.  We
have studied the properties of several types of ETC models in our work.  One of
these has a gauge group $G = {\rm SU}(5)_{ETC} \times {\rm SU}(2)_{HC} \times
G_{SM}$.  In addition to ETC, this has another gauge interaction, hypercolor
(HC), which helps to produce the desired ETC symmetry-breaking pattern.  The SM
fermions and corresponding technifermions transform according to the
representations
\beq
Q_L: \ (5,1,3,2)_{1/3,L} \ , \quad \  u_R: \ (5,1,3,1)_{4/3,R} \ ,
                             \quad \  d_R: \ (5,1,3,1)_{-2/3,R} \ ,
\label{qreps}
\eeq
\beq
L_L: \ (5,1,1,2)_{-1,L}, \quad\quad  e_R: \ (5,1,1,1)_{-2,R} \ ,
\eeq
where the subscripts denote $Y$. For example, writing out the components of
$e_R$, one has $(e^1, e^2, e^3, e^4, e^5)_R \ \equiv \ (e, \mu, \tau, E^4,
E^5)_R$, where the last two entries are the charged technileptons.  In
addition, the model includes the SM-singlet ETC-nonsinglet fermions
\beq
\psi_{ij,R}: \ (\overline{10},1,1,1)_{0,R} \ , \quad\quad
 \zeta^{ij,\alpha}_R: \ (10,2,1,1)_{0,R} \ , \quad\quad
\omega^{\alpha}_{p,R} : \ 2(1,2,1,1)_{0,R}
\eeq
where here $1 \le i,j \le 5$ are SU(5)$_{ETC}$ indices, $\alpha=1,2$ is an
SU(2)$_{HC}$ index, and $p=1,2$ is a copy number.  
In this model the SM fermions and corresponding technifermions have vectorial
couplings to ETC gauge bosons, but the SM-singlet sector makes the full ETC 
theory a chiral gauge theory.  Hence, there are no fermion mass terms in 
the lagrangian.  This theory has no gauge or global anomalies.

Since the ETC theory is asymptotically free, its gauge coupling grows as the
energy scale decreases.  The ETC breaking occurs because of the formation of
ETC-noninvariant bilinear condensates of ETC-nonsinglet, SM-singlet fermions.
To analyze the stages of symmetry breaking, we identify plausible preferred
condensation channels using a generalized most-attractive-channel (MAC) 
approach.  We envision that as the energy decreases from high values down to $E
\sim \Lambda_1 \sim 10^3$ TeV, the coupling $\alpha_{_{ETC}}$ becomes 
sufficiently large to produce condensation in the attractive channel
$(\overline{10},1,1,1)_{0,R} \times (\overline{10},1,1,1)_{0,R} \to
(5,1,1,1)_0$, breaking ${\rm SU}(5)_{ETC} \to {\rm SU}(4)_{ETC}$.  With respect
to the unbroken ${\rm SU}(4)_{ETC}$, we have $(\overline{10},1,1,1)_{0,R} =
(\bar 4,1,1,1)_{0,R} + (\bar 6,1,1,1)_{0,R}$; we denote the $(\bar
4,1,1,1)_{0,R}$ as $\alpha_{1i,R} \equiv \psi_{1i,R}$ for $2 \le i \le 5$ and
the $(\bar 6,1,1,1)_{0,R}$ as $\xi_{ij,R} \equiv \psi_{ij,R}$ for $2 \le i,j
\le 5$.  The associated condensate is then
\beq
\langle \epsilon^{1 i j k \ell} \xi^T_{ij,R} C \xi_{k \ell,R} \rangle =
8 \langle \xi^T_{23,R} C \xi_{45,R} - \xi^T_{24,R} C \xi_{35,R} +
\xi^T_{25,R} C \xi_{34,R} \rangle \ . 
\eeq
The six fields $\xi_{ij,R}$, $2 \le i,j \le 5$, involved in this condensate
gain dynamical masses $\simeq \Lambda_1$. 

At energy scales below $\Lambda_1$, depending on relative strengths of gauge
couplings, different symmetry-breaking sequences can occur.  Again, these arise
via the dynamical formation of condensates involving SM-singlet ETC-nonsinglet
fermions.  For example, one sequence, S1, leads to the breaking 
${\rm SU}(4)_{ETC} \to {\rm SU}(3)_{ETC}$ at $\Lambda_2 \simeq 50-100$ TeV and
finally ${\rm SU}(3)_{ETC} \to {\rm SU}(2)_{TC}$ at $\Lambda_3 \simeq$ few TeV,
with SU(2)$_{HC}$ unbroken.  Another sequence, S2, involves a breaking 
${\rm SU}(4)_{ETC} \to {\rm SU}(2)_{ETC}$ at $\Lambda_{23} \simeq 50$ TeV and
${\rm SU}(2)_{HC} \to {\rm U}(1)_{HC}$ at a scale $\Lambda_{BHC} \lsim
\Lambda_1$. In all cases, SU(2)$_{TC}$ is an exact gauge symmetry. 
At the lowest scale, $\Lambda_{TC}$, the technifermion condensates form,
breaking electroweak symmetry and giving masses to the $W$ and $Z$.

\begin{figure}[hbtp]
\begin{center}
\includegraphics[3in, 8in][4in, 9.5in]{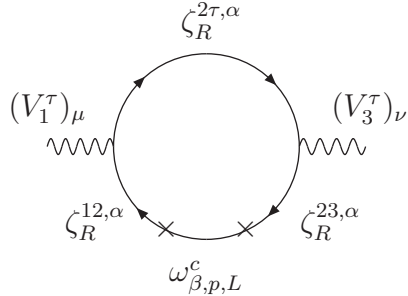}
\end{center}
\vspace{-1mm}
\caption{Graph for $V^\tau_1 \to V^\tau_3$ with $\tau=4,5$, for sequence S1.} 
\label{ckmfig10}
\end{figure}

\begin{figure}[hbtp]
\begin{center}
\includegraphics[3in, 8in][4in, 9.5in]{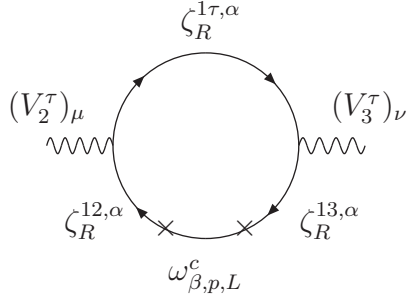}
\end{center}
\vspace{-1mm}
\caption{Graph for $V^\tau_2 \to V^\tau_3$ with $\tau=4,5$, for sequence 
S2.} 
\label{ckmfig14}
\end{figure}

Certain fermion condensates contribute to nondiagonal ETC gauge boson
propagator corrections and hence mixing.  For example, in sequence $S_1$, the
mixing $V^\tau_1 \to V^\tau_3$, $\tau =4,5$ is induced by the graph in Fig. 
\ref{ckmfig10}, 
with the result
\beq
{}^\tau_3 \Pi^\tau_1(0) \simeq \frac{\alpha_{_{ETC}}}{4\pi} \, \int \,
(k^2 dk^2) \frac{k^4 \, \Sigma_3(k)^2}{[k^2 + \Sigma_3(k)^2]^4} \ , 
\eeq
where $\Sigma_3 \simeq \Lambda_3$.  This yields
${}^\tau_3 \Pi^\tau_1(0) \simeq {\rm const.} \times \Lambda_3^2$.
In sequence $S_2$, the mixing
$V^\tau_2 \to V^\tau_3$, $\tau =4,5$ is induced by the graph in Fig. 
\ref{ckmfig14}, 
giving ${}^\tau_3 \Pi^\tau_2(0) \simeq {\rm const.} \times \Lambda_{23}^2$.
We find that the feature that nondiagonal ETC gauge boson propagator
corrections ${}^\tau_i \Pi^\tau_j(0)$ are proportional to the square of the
lowest ETC scale (or smaller) is generic in this type of ETC model, reflecting
a type of approximate generational symmetry.  Other ETC gauge boson mixings are
similarly suppressed.

\section{Fermion Masses and Mixing}

\begin{figure}[hbtp]
\begin{center}
\includegraphics[3in, 8in][4in, 9.5in]{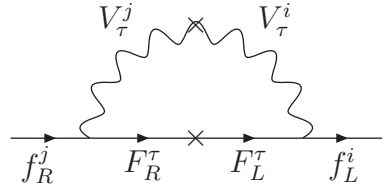}
\end{center}
\vspace{-1mm}
\caption{Graph contributing to the fermion mass matrix element 
$M^{(f)}_{ij}$.} 
\label{mfij}
\end{figure}

Figure \ref{mfij} shows a one-loop graph contributing to the mass matrix
element $M^{(f)}_{ij}$ for a SM fermion $f$ (up- or down-type quark or charged
lepton) appearing in the operator $\bar f_{i,L} M^{(f)}_{ij} f_{j,R} + h.c.$.
In this figure we distinguish the first three ETC indices, which refer to SM
fermion generations, and the indices 4,5 which are TC, by denoting the latter
as $\tau=4,5$.
An estimate for diagonal entries is (with no sum on $i$)
\beq
M^{(f)}_{ii} \simeq \frac{\kappa \, (N_{TC}/2) \, \eta \,
\Lambda_{TC}^3} {\Lambda_i^2} \ , 
\eeq
where $\kappa \simeq O(10)$ is a numerical factor from the integral and in WTC,
$\eta \simeq \Lambda_3/\Lambda_{TC}$.  This is only a rough estimate, since
ETC coupling is strong, so higher-order diagrams are also important.
The sequential breaking of the ETC symmetry at the highest scale $\Lambda_1$,
the intermediate scale $\Lambda_2$, and the lowest scale $\Lambda_3$, thus
produces the generational hierarchy in the fermion masses.

Insertions of the nondiagonal ETC propagator corrections (indicated by the
cross on the gauge boson line in Fig. \ref{mfij}) give rise to 
off-diagonal elements of the $M^{(f)}$. These have the form 
\beq
M^{(f)}_{ij} \simeq \frac{\kappa \, \eta \, \Lambda_{TC}^3 \, {}^j_\tau
\Pi^i_\tau}{\Lambda_i^2 \Lambda_j^2} \ . 
\eeq
Although the resultant mixing is not fully realistic, this model shows how not
just diagonal but also off-diagonal elements of SM fermion mass matrices
$M^{(f)}$ and hence CKM mixing, could arise dynamically \cite{ckm}.  Further
corrections to $M^{(f)}$ arise from SM gauge interactions, in particular,
SU(3)$_c$ and U(1)$_Y$, and from direct diagonal ETC gauge boson exchanges, in
particular, $V_{d3}$ (having mass $\sim \Lambda_3$ and corresponding to the 
SU(5)$_{ETC}$ generator $T_{d3}=(2\sqrt{3})^{-1}{\rm diag}(0,0,-2,1,1)$).

\section{Mechanism for Light Neutrinos}

 An old puzzle in TC/ETC theories was how to explain light neutrino masses.  A
solution to this was given in \cite{nt} and analyzed further in \cite{lrs,ckm}.
The $\alpha_{1j,R}$ with $j=2,3$ are right-handed electroweak-singlet neutrinos
and get induced Dirac neutrino mass terms connecting with $(n^1,n^2,n^3)_L =
(\nu_e,\nu_\mu,\nu_\tau)_L$.  These Dirac masses $\bar n_{i,L} M_D
\alpha_{1j,R}+ h.c.$ cannot be generated by the usual one-loop ETC graphs that
produce diagonal quark and charged lepton masses and are thus suppressed.  The
$\alpha_{1j,R}$ also have induced Majorana mass terms $\alpha_{1i,R}^T C r_{ij}
\alpha_{1j,R}+h.c.$ For example, with sequence S2, denoting the relevant Dirac
mass terms as $b_{ij}$, one finds $b_{ij} \simeq \kappa \Lambda_{TC}^4/
\Lambda_{23}^3 \simeq O(0.1)$ MeV for $2 \le i,j \le 3$; further, $r_{23}
\simeq \kappa \Lambda_{BHC}^3 \Lambda_{23}^3/\Lambda_1^5$.  Numerically,
$|r_{23}| \simeq O(10^2)$ GeV.  The resultant electroweak-nonsinglet neutrinos
are, to very good approximation, linear combinations of three mass eigenstates,
with $\nu_3$ mass
\beq
m(\nu_3) \simeq \frac{(b_{23}+b_{22})^2}{r_{23}} \simeq
\frac{\kappa \Lambda_{TC}^8 \Lambda_1^5}{\Lambda_{23}^9 \Lambda_{BHC}^3} \ . 
\label{mnu3}
\eeq
With the above-mentioned numerical values and
$\Lambda_{BHC} \lsim \Lambda_1$, we find $m(\nu_3) \simeq 0.05$ eV,
consistent with experimental results on neutrino oscillations.  
Similarly, 
\beq
\frac{m(\nu_2)}{m(\nu_3)} =
\bigg ( \frac{b_{23}-b_{22}}{b_{23}+b_{22}} \bigg )^2 \ , 
\label{mnu2}
\eeq
which is again consistent with experimental data.  Since $|r_{23}| >>
|b_{ij}|$, this is a seesaw, but quite different from the SUSY GUT seesaw; the
Majorana masses $r_{ij}$ that underly the seesaw are not GUT-scale and are
actually much smaller than the ETC scales $\Lambda_i$.

\section{Constraints from Neutral Flavor-Changing Current Processes}

 An early concern was that ETC interactions would lead to excessively large
flavor-changing neutral current processes.  A particularly severe constraint
arises from $K^0 - \bar K^0$ mixing.  Early treatments wrote the effective
Lagrangian for this as
\beq
{\cal L}_{eff} \simeq \frac{1}{\Lambda_{ETC}^2} [\bar s \gamma_\mu d]^2 \ , 
\eeq
where $\Lambda_{ETC}$ was a generic ETC scale.  To suppress FCNC effects
adequately, it was thought that $\Lambda_{ETC}$ had to be so high that there
would be excessive suppression of SM fermion masses.  However, we have shown,
using our reasonably UV-complete theories, that in the present type of ETC
theory this old view was too pessimistic; ETC contributions to FCNC processes
are smaller than had been inferred with the above naive ${\cal L}_{eff}$,
because of residual approximate generational symmetries \cite{ckm}-\cite{kt}.

The coupling of the ETC gauge bosons to the SM fermion mass eigenstates is
given by 
\beq
{\cal L}_{int} = g_{_{ETC}} \sum_{f,j,k}
\bar f_j \gamma_\lambda ({\cal V}^\lambda)^j_k f^k
\equiv g_{_{ETC}} \sum_{f,j,k}
\bar f_{m,j} \gamma_\lambda (A^\lambda)^j_k  f^k_m \ , 
\eeq
where ${\cal V}$ is a matrix containing the ETC gauge fields, $A^\lambda \equiv
U^{(f)} {\cal V}^\lambda U^{(f) \ -1}$, and the $U^{(f)}$ are the unitary
transformations that diagonalize $M^{(f)}$ via $U^{(f)} M^{(f)} U^{(f) \ -1} =
M^{(f)}_{diag.}$. We parametrize each $U^{(f)}$ with a PDG-type parametrization
in terms of $\theta_{12}^{(f)}$, $\theta_{13}^{(f)}$, $\theta_{23}^{(f)}$, and
$\delta^{(f)}$.  The approximate generational symmetry in the ETC sector can
naturally yield relatively small CKM mixing with $\theta^{(f)}_{jk}$ depending
on ratios of smaller to larger ETC scales.

We have analyzed ETC contributions to a number of processes, including the
neutral meson mixings $K^0 - \bar K^0$, $D^0 - \bar D^0$, $B^0_d - \bar B_d$,
$B^0_s - \bar B_s$, and decays such as $\mu^+ \to e^+ e^+ e^-$, involving
four-fermion operators \cite{ckm,kt}.  As an example, consider $K^0 - \bar K^0$
mixing and the resultant $K_L - K_S$ mass difference $\Delta m_{K_L K_S}$.  The
SM contribution is consistent with the experimental value, $\Delta m_{K_L
K_S}/m_K = 0.70 \times 10^{-14}$. A key to the suppression is the fact that, in
terms of ETC eigenstates, an $s \bar d$ in a $\bar K^0$ produces a $V^2_1$ ETC
gauge boson, but this cannot directly yield a $d \bar s$ in the final-state
$K^0$; the latter is produced by a $V^1_2$.  Hence, this requires either the
ETC gauge boson mixing $V^2_1 \to V^1_2$ or the mixing of ETC quark eigenstates
to produce mass eigenstates.  The contribution from $V^2_1 \to V^1_2$ yields a
coefficient
\beq
c \ \sim \ \frac{1}{\Lambda_1^2} \ {}^1_2 \Pi^2_1 \ \frac{1}{\Lambda_1^2}
\ \sim \ \frac{\Lambda_3^2}{\Lambda_1^2}  \frac{1}{\Lambda_1^2} \ \ll \ 
\frac{1}{\Lambda_1^2} \ . 
\label{ccalc} 
\eeq
With above values, $\Lambda_1 \sim 10^3$ TeV, $\Lambda_3 \sim 3$ TeV, the
suppression factor is $(\Lambda_3/\Lambda_1)^2 \simeq 10^{-5}$.  So, rather
than the naive result $\Delta m_{K_L K_S}/m_K \sim
\Lambda_{QCD}^2/\Lambda_1^2$, this yields the much smaller result $\Delta
m_{K_L K_S}/m_K \sim \Lambda_3^2 \, \Lambda_{QCD}^2/\Lambda_1^4 \sim 10^{-18}$.
Hence, the dominant ETC contributions arise from the mixing of ETC eigenstates
of quarks to form mass eigenstates.  First, $s \bar d$ can couple to $V_{d2}$
(having mass $\simeq \Lambda_2$ and corresponding to the SU(5)$_{ETC}$
generator $T_{d2}=(2\sqrt{6})^{-1}{\rm diag}(0,-3,1,1,1)$), with coefficient
$\theta^{(d)}_{12}$.  The resultant diagram with exchange of this gauge boson
$V_{d2}$ contributes $\sim (\theta^{(d)}_{12})^2/\Lambda_2^2$ to the amplitude.
Requiring this to be small relative to SM contribution yields the constraint
$|\theta_{12}^{(d)}| \lsim 10^{-2}$. Second, $s \bar d$ can couple to $V_{d3}$,
with coefficient $\theta^{(d)}_{13}\theta^{(d)}_{23}$.  The resultant diagram
contributes $\sim (\theta^{(d)}_{13}\theta^{(d)}_{23})^2/ \Lambda_3^2$ to the
amplitude.  This yields the constraint $|\theta_{13}^{(d)}\theta_{23}^{(d)}|
\lsim 0.4 \times 10^{-3}$.  A comprehensive analysis of constraints from
processes involving dimension-six four-fermion operators was given in
\cite{ckm,kt}.  A related analysis is \cite{lm}.

We have also studied ETC contributions to (dimension-5) diagonal and transition
lepton and quark dipole moments, and constraints from limits on $\mu^+ \to e^+
\gamma$, $\tau^+ \to \ell^+ \gamma$, the measured muon $(g-2)$ and $b \to s
\gamma$ decay, and limits on lepton, neutron, atomic electric dipole moments
\cite{dml}.  Again, we found that this type of ETC model can be consistent
with existing data. 

\section{Precision Electroweak Constraints}

One may ask what the momentum dependence of a SM fermion mass is in this type
of theory.  This question was answered in \cite{sml}; 
the running mass $m_{f_j}(p)$ exhibits the power-law decay
\beq
m_{f_j}(p) \propto \frac{\Lambda_j^2}{p^2}
\label{mfp}
\eeq
for Euclidean momenta $p \gg \Lambda_j$, where $f_j$ is a fermion of generation
$j$. (Here we neglect logarithmic factors, which are subdominant relative to
this power-law falloff.)  Thus, $m_t(p)$ and $m_b(p)$ decay like
$\Lambda_3^2/p^2$ for $p \gg \Lambda_3$, while $m_u(p)$ and $m_d(p)$ are hard
up to the much higher scale $\Lambda_1$.  We have investigated whether
precision electroweak data are consistent with these power-law decays of SM
fermion masses and have found that they are.  The largest effects occur for
$m_t$.  Consider, e.g., the $t$-quark contribution to $\rho$.  The conventional
(hard, one-loop) result for this is $(\Delta \rho )_{t,hard} \simeq 3 G_f
m_t^2/(8 \pi^2 \sqrt{2})$.  The power-law decay of $m_t$ above $\Lambda_3$
changes this to $(\Delta \rho)_{t} = (\Delta \rho)_{t,hard}[ 1 - a_\rho
(m_t^2/\Lambda_3^2)]$, where $a_\rho$ is positive, $\sim O(1)$.  The softness
of $m_t$ thus slightly reduces the violation of custodial symmetry.  We find a
similarly small change in the $(t,b)$ contribution to $S$ relative to the
conventional hard-mass result.

As noted, the $S$ parameter places a stringent constraint on TC/ETC theories.
A naive perturbative calculation gives, for the TC contribution to $S$, the
result $S_{pert.} \simeq N_{TC} \, N_D/(6\pi)$, where $N_D$ denotes
the number of technifermion EW doublets (given that the dynamical masses of the
technifermions in each EW doublet are nearly degenerate, as should be true to
satisfy the $\rho$-parameter constraint).  However, this calculation assumes
that the technifermions are weakly interacting, whereas actually, they are
strongly interacting, at the relevant scale, $\sim m_Z$; hence, this
perturbative formula cannot be expected to be reliable.  This is
important, since otherwise, even with the minimal value, $N_{TC}=2$, it would
yield an excessively large value of $S_{pert.} = 4/(3\pi) \simeq 0.4$
for a one-family TC model (and the marginally acceptable value of 
$S_{pert.} = 1/(3\pi) \simeq 0.1$ for a TC model with one EW technifermion
doublet.  Experimentally, $S \lsim 0.1$.  Several studies have found that
walking can reduce $S$ \cite{scalc}-\cite{ads},\cite{s}.  In
particular, Refs. \cite{hky} and \cite{s} have used solutions of
Schwinger-Dyson and Bethe-Salpeter equations to evaluate $S$ \cite{mk}.

One way to reduce $S$ would be to reconsider TC models with only one
technifermion electroweak doublet.  But with this technifermion content alone,
these models would not have the walking behavior that is needed not just to
reduce $S$ but also to generate adequate fermion masses.  One can maintain the
necessary walking behavior in a TC model with one EW doublet of technifermions
by adding a requisite set of SM-singlet technifermions \cite{ts}.  The
SU(2)$_{TC}$ theory should have $N_f \simeq 8$ Dirac technifermions for
walking.  The electroweak-nonsinglet technifermions comprise two, so we need
six more.  One adds six SM-singlet, Dirac fermions (i.e., 12 chiral components)
that transform as 2's under SU(2)$_{TC}$.  These do not contribute to $S$ or
$T$. Note that in this theory, $[G_{ETC}, G_{SM}] \ne 0$, so the ETC gauge
bosons carry color and charge.  Embedding TC in ETC is thus more complicated
than for one-family TC models.  Another approach is to use technifermions in
higher-dimensional representations \cite{lanerep,fs}, \cite{ts}. In general,
the $S$ constraint remains a concern for TC/ETC theories.

\section{Splitting of $t$ and $b$ Masses}

Another challenge for TC/ETC models is to account for the splitting of the $t$
and $b$ masses without excessive contributions to $\rho-1$ (violation of
custodial SU(2) symmetry).  One cannot do this by having $\Sigma_{TC,U}$
significantly larger than $\Sigma_{TC,D}$ since this would violate the
custodial symmetry too much.  One could consider trying to achieve this
splitting using a class of ETC models in which left and right components of
up-type quarks and techniquarks transform the same way under SU(5)$_{ETC}$, but
the left and right components of down-type quarks and techniquarks transform
according to relatively conjugate representations.  However, we showed that
such models have serious problems with excessively large FCNC's \cite{kt}.  For
example, consider a model in which $Q_L$ and $u_R$ are 5's of SU(5)$_{ETC}$ and
$d_R$ is a $\bar 5$.  So, e.g., the $\bar K^0 - K^0$ transition can proceed
directly via $s_L \bar d_L \to V^2_1 \to d_R \bar s_R$ without ETC gauge boson
mixing.  This gives too large a value for $\Delta m_{K_L K_S}$.

A different idea would be to use two ETC gauge groups, say ${\rm
SU}(5)_{ETC} \times {\rm SU}(5)_{ETC}'$ such that the left- and right-handed
components of charge $Q=2/3$ quarks transform under the same ETC group, while
left- and right-handed components of charge $-1/3$ quarks and charged leptons
transform under different ETC groups \cite{aes,met}.
These models thereby suppress the masses $m_b$ and $m_\tau$ relative to $m_t$,
etc. because generating $m_b$ requires mixing between the two ETC groups, which
is suppressed, while $m_t$ does not.  However, they tend to produce too much 
suppression of the masses of first- and second-generation down-type quarks and
charged leptons \cite{met}.  Thus, a satisfactory explanation of $t-b$ 
splitting appears to remain a challenge for ETC models.

\section{Dynamical Breaking of Higher Gauge Symmetries} 

To what extent can one embed a TC/ETC in a theory having higher gauge symmetry,
using dynamical symmetry breaking to break this higher symmetry?  Such higher
unification would be desirable in order to explain features not explained by
the standard model, including charge quantization, prediction of relative sizes
of gauge couplings, and unification of quarks and leptons.  In \cite{lrs} 
we constructed asymptotically free gauge theories exhibiting dynamical
breaking of the left-right, strong-electroweak gauge group
\beq
G_{LR} = {\rm SU}(3)_c \times {\rm SU}(2)_L \times {\rm SU}(2)_R \times
{\rm U}(1)_{B-L}
\label{glr}
\eeq
(where $B$ and $L$ denote baryon and lepton number) and 
its extension to 
\beq
G_{422}={\rm SU}(4)_{PS} \times {\rm SU}(2)_L \times {\rm SU}(2)_R \ , 
\label{g422}
\eeq
where ${\rm SU}(4)_{PS} \supset {\rm SU}(3)_c \times {\rm U}(1)_{B-L}$ is the
Pati-Salam group.  These models technicolor for electroweak breaking, and
extended technicolor for the breaking of $G_{LR}$ and $G_{422}$ and the
generation of fermion masses, including a seesaw mechanism for neutrino masses.
These models explain why $G_{LR}$ and $G_{422}$ break to ${\rm SU}(3)_c \times
{\rm SU}(2)_L \times {\rm U}(1)_Y$, and why this takes place at a scale ($\sim
10^3$ TeV) which is large compared to the electroweak scale.  In particular,
the model with the gauge group $G_{422}$ achieves charge quantization in the
context of dynamical breaking of all symmetries.

We have also investigated various more ambitious unification schemes involving
TC/ETC \cite{tg}.  In particular, we have studied the possibility of unifying a
one-family technicolor group with the SM gauge group in a simple group, but
find that it appears difficult to obtain the requisite symmetry breaking
dynamically.

\section{Other Phenomenology}

There are several other relevant topics for discussion.  One has to do with the
spectrum of these theories.  As a confining gauge theory, technicolor produces
a spectrum of TC-singlet techni-hadrons composed of technifermions and
technigluons.  This spectrum exhibits differences in a walking theory, as
compared with a QCD-like theory.  For example, while $m_{a_1}/m_\rho = 1.6$ in
QCD, the analogous ratio $m_{(a_1)_{TC}}/m_{\rho_{TC}}$ is close to unity in
the walking limit \cite{mm}.  This is natural, since a theory with walking
behavior has $N_f$ near to, although slightly less than, the critical value
$N_{f,cr}$ at which the theory would go over from the confined phase with
spontaneous chiral symmetry breaking to a chirally symmetric phase.  It is of
interest to investigate how these and other hadron masses change as one
decreases $N_f$ (hence increases $\alpha_{IR}$) to move from the walking regime
near $N_{f,cr}$ toward the QCD-like regime at smaller $N_f$; this was done in
\cite{sg}. Searches for the production and decays of these technihadrons at the
CERN Large Hadron Collider will be of great importance in testing technicolor
theories.  For example, by analogy with $\rho \to \pi \pi$ in QCD, since the
technipions are absorbed to become the longitudinal components of the $W$ and
$Z$, one would have $\rho_{TC}^0 \to W_L^+ W_L^-$, $\rho_{TC}^+ \to W_L^+
Z_L^0$, etc.  A related issue concerns pseudo-Nambu Goldstone bosons.  While
walking raises the masses of many of these PNGB's, they can present a
phenomenological challenge for the model. It will be natural to search further
for these at the LHC. 

\section{Summary}

The question of the origin of electroweak symmetry is still not answered, and
dynamical EWSB via technicolor and extended technicolor remains an 
interesting possibility.  This approach is strongly constrained by
flavor-changing neutral current data and precision electroweak measurements. 
We have reviewed here some recent progress on TC/ETC models.  Clearly, there
are a number of challenges for such TC/ETC models.  Soon, experiments at the 
LHC will show whether these ideas are realized in nature.

\section*{Acknowledgments}
I would like to thank my collaborators on works discussed here, T. Appelquist
and M. Piai and, in subsequent papers, N. Christensen and M. Kurachi.  I also
thank M. Harada, M. Tanabashi, and K. Yamawaki, for organizing this SCGT06
conference.  Our research was partially supported by the grant
NSF-PHY-03-54776.


\begin{thebibliography}{40}

\bibitem{tc}
S. Weinberg, Phys. Rev. D {\bf 19}, 1277 (1979);
L. Susskind, Phys. Rev. D {\bf 20}, 2619 (1979); see also 
S. Weinberg, Phys. Rev. D {\bf 13}, 974 (1976).

\bibitem{etc}
S. Dimopoulos and L. Susskind, Nucl. Phys. B {\bf 155}, 237 (1979);
E. Eichten and K. Lane, Phys.  Lett. B {\bf 90}, 125 (1980).

\bibitem{etcrev}
Some recent reviews are K. Lane, hep-ph/0202255; C. Hill and E. Simmons,
Phys. Rep. {\bf 381}, 235 (2003);
R. S. Chivukula, M. Narain, J. Womersley, in http://pdg.lbl.gov

\bibitem{nt}
T. Appelquist and R. Shrock, Phys. Lett. {\bf B548}, 204 (2002).

\bibitem{lrs}
T. Appelquist and R. Shrock, Phys. Rev. Lett. {\bf 90}, 201801 (2003).

\bibitem{ckm}
T. Appelquist, M. Piai, and R. Shrock, Phys. Rev. D {\bf 69}, 015002 (2004).

\bibitem{dml}
T. Appelquist, M. Piai, and R. Shrock, Phys. Lett. B {\bf 593}, 175 (2004); 
{\it ibid.}, {\bf 595}, 442 (2004).

\bibitem{kt}
T. Appelquist, N. Christensen, M. Piai, and R. Shrock, Phys. Rev. D {\bf 70},
093010 (2004).

\bibitem{sml}
N. Christensen and R. Shrock, Phys. Rev. Lett. {\bf 94}, 241801 (2005).

\bibitem{tg}
N. Christensen and R. Shrock,  Phys. Rev. D {\bf 72}, 035013 (2005).

\bibitem{ts}
N. Christensen and R. Shrock,  Phys. Lett. B {\bf 632}, 92 (2006).

\bibitem{met}
N. Christensen and R. Shrock,  Phys. Rev. D {\bf 74}, 015004 (2006).

\bibitem{s}
M. Kurachi and R. Shrock, Phys. Rev. D {\bf 74}, 056003 (2006).

\bibitem{sg}
M. Kurachi and R. Shrock, JHEP {\bf 12}, 034 (2006).

\bibitem{mk}
M. Kurachi, in these SCGT06 proceedings. 

\bibitem{ssvz}
P. Sikivie, L. Susskind, M. Voloshin, and V. Zakharov, Nucl. Phys. B 
{\bf 173}, 189 (1980). 

\bibitem{onefamily}
See, e.g., E. Farhi and L. Susskind, Phys. Rept. {\bf74}, 277 (1981). 

\bibitem{wtc1}
B. Holdom, Phys. Lett. B {\bf 150}, 301 (1985).

\bibitem{wtc2} 
K. Yamawaki, M. Bando, and K. Matumoto, Phys. Rev. Lett. {\bf
56}, 1335 (1986).

\bibitem{chipt1}
T. Appelquist, D. Karabali, and L. C. R. Wijewardhana, Phys. Rev. Lett. {\bf
57}, 957 (1986); T. Appelquist and L. C. R. Wijewardhana, Phys. Rev. D
{\bf 35}, 774 (1987); Phys. Rev. D {\bf 36}, 568 (1987).

\bibitem{at94}
T. Appelquist and J. Terning, Phys. Rev. D {\bf 50}, 2116 (1994).

\bibitem{chipt2}
T. Appelquist, J. Terning, and L. C. R. Wijewardhana,
Phys. Rev. Lett.  {\bf 77}, 1214 (1996); 
T. Appelquist, A. Ratnaweera, J. Terning, and 
L. C. R. Wijewardhana, Phys. Rev. D {\bf 58}, 105017 (1998).

\bibitem{my}
V. Miransky and K. Yamawaki, Phys. Rev. D {\bf 55}, 5051 (1997).

\bibitem{chiv}
R. S. Chivukula, Phys. Rev. D {\bf 55}, 5238 (1997).

\bibitem{pt}
M. Peskin and T. Takeuchi, Phys. Rev. Lett.  {\bf 65}, 964 (1990);
Phys. Rev. D {\bf 46}, 381 (1992).

\bibitem{lm}
K. Lane and A. Martin, Phys. Rev. D {\bf 71}, 015011 (2005). 

\bibitem{scalc}
R. Cahn and M. Suzuki, Phys. Rev. D {\bf 44}, 3641 (1991);
T. Appelquist and G. Triantaphyllou, Phys. Lett. B {\bf 278}, 345 (1992);
R. Sundrum and S. Hsu, Nucl. Phys. B {\bf 391}, 127 (1993);
T. Appelquist and F. Sannino, Phys. Rev. D {\bf 59}, 067702 (1999);
S. Ignjatovic, L. C. R. Wijewardhana, and T. Takeuchi, Phys. Rev.
D {\bf 61}, 056006 (2000).

\bibitem{hky}
M. Harada, M. Kurachi, and K. Yamawaki, Prog. Theor. Phys. {\bf 115}, 765 
(2006). 

\bibitem{ads}
D. K. Hong and H.-U. Yee, Phys. Rev. D {\bf 74}, 015011 (2006); 
J. Hirn and V. Sanz, Phys. Rev. Lett. {\bf 97}, 121803 (2006); 
M. Piai, hep-ph/0608241. 

\bibitem{lanerep}
E. Eichten and K. Lane, Phys. Lett. B {\bf 222}, 274 (1988). 

\bibitem{fs}
D. Hong, S. Hsu, and F. Sannino, Phys. Lett. B {\bf 597}, 89 (2004); 
F. Sannino and K. Tuominen, Phys. Rev. D {\bf 71}, 051901 (2005).

\bibitem{aes}
T. Appelquist, N. Evans, and S. Selipsky, Phys. Lett. B {\bf 374}, 145 (1996).

\bibitem{mm}
M. Harada, M. Kurachi, and K. Yamawaki, Phys. Rev. D {\bf 68}, 076001 (2003).

\end{thebibliography}
\end{document}